\documentclass[prl,aps,twocolumn,showpacs]{revtex4}

\usepackage{graphicx}

\begin{document}

\title{Classical and Quantum Fluctuation Theorems for Heat Exchange}
\author{Christopher Jarzynski}
\affiliation{Theoretical Division, T-13, MS B213,
  Los Alamos National Laboratory,
  Los Alamos, NM 87545}
\email{chrisj@lanl.gov}
\author{Daniel K. W\'ojcik}
\affiliation{Center for Nonlinear Science, School of Physics, Georgia
  Institute of Technology, 837 State Street, Atlanta, GA 30332-0430,
  USA} 
\affiliation{Department   of  Neurophysiology,  Nencki   Institute  of
  Experimental Biology, 3 Pasteur Str., 02-093 Warsaw, Poland}
\email{danek@cns.physics.gatech.edu} 
\date{\today}

\begin{abstract}
  The  statistics of heat  exchange between  two classical  or quantum
  finite  systems  initially prepared  at  different temperatures  are
  shown to obey a fluctuation theorem.
\end{abstract}

\pacs{05.70.Ln, 05.20.-y \\
      Keywords: {\bf fluctuation theorem},
                {\bf irreversible processes}
      %LAUR-99-5580
     }
\maketitle

The  {\it  fluctuation  theorem}   (FT)  refers  to  a  collection  of
theoretical   predictions~\cite{EvansCM93,   EvansS94,  GallavottiC95,
  Kurchan98,   LebowitzS99,  Maes99,   Evans02},   recently  confirmed
experimentally~\cite{WangSMSE02},  pertaining  to  a  system  evolving
under   non-equilibrium  conditions.    These   results  are   roughly
summarized by the equation
\begin{equation}
  \label{eq:ft}
  \ln \frac{p(+\Sigma)}{p(-\Sigma)} = \Sigma,
\end{equation}
where $p(\Sigma)$  denotes the probability  that an amount  of entropy
$\Sigma$  is generated during  a specified  time interval.   Both {\em
  transient\/} and {\em  steady state\/} versions of the  FT have been
obtained.  The definition  of ``entropy generated'' ($\Sigma$) depends
on  the dynamics  used  to model  the  evolution of  the system  under
consideration.  However,  for a variety of physical  situations, and a
variety  of equations  of motion  (both deterministic  and stochastic)
used  to model  them, the  FT  has been  established under  reasonable
definitions   of   entropy    generation.    Moreover,   the   FT   is
related~\cite{crooks99} to a set of {\em free energy relations\/} (see
e.g.~\cite{Jarzynski97,Crooks98})   which  connect   equilibrium  free
energy  differences to  non-equilibrium  work values,  and which  have
recently been confirmed experimentally~\cite{Liphardt02}.

The    situations   modeled   in    Refs.~\cite{EvansCM93,   EvansS94,
  GallavottiC95,  Kurchan98, LebowitzS99,  Maes99,  crooks99, Evans02,
  WangSMSE02, Jarzynski97, Crooks98}  all involve an externally driven
system,  in the presence  of a  heat reservoir.   The purpose  of this
paper  is to  point out  that a  similar result  can be  derived  in a
different  setting.   Namely,  we  will  obtain  a  symmetry  relation
constraining  the  statistics  of  heat exchange  between  two  bodies
initially prepared  at different  temperatures.  We will  present both
classical and quantum derivations, and will use the term {\it exchange
  fluctuation theorem} (XFT) to refer to these results.

In what  follows, the XFT  (Eq.~(\ref{eq:heatft})) will be  stated and
derived.    A  corollary  result   related  to   the  Second   Law  of
Thermodynamics will then be presented (Eq.~(\ref{eq:bound})).

Consider  two  finite bodies,  $A$  and  $B$,  separately prepared  in
equilibrium states at temperatures $T_A$ and $T_B$, respectively, then
placed in thermal contact with one another for a time $\tau$, and then
separated again.  Let $Q$ denote the net heat transfer from $A$ to $B$
during the interval of contact, i.e.  the amount of energy lost by $A$
and gained by $B$.  Now  imagine repeating this experiment many times,
always initializing the two  bodies at the specified temperatures, and
let $p_\tau(Q)$ denote the observed distribution of values of $Q$ over
the  ensemble of repetitions.   Then we  claim that  this distribution
satisfies
\begin{equation}
  \label{eq:heatft}
  \ln
  {p_\tau(+Q)\over p_\tau(-Q)} 
  = \Delta\beta \cdot Q,
\end{equation}
where  $\Delta\beta=T_B^{-1}-T_A^{-1}$ is  the difference  between the
inverse temperatures at which the bodies are prepared.

In  the  quantum case  we  must  define  $Q$ through  an  experimental
procedure:  starting  with  the  two  systems  initially  prepared  at
different temperatures,  we first measure  the energy of  each system,
then we allow them to weakly  interact over a time $\tau$, and finally
we again  measure the energy of  each system.  We  then interpret heat
transfer  in   terms  of  the  changes  in   these  measured  energies
(Eq.~(\ref{eq:12})). This approach is  similar in spirit to that taken
by~\cite{kurchan00s,  Mukamel03,  monnai03},  who  considered  related
problems.  For an alternative approach see e.g.~\cite{deroeck03}.

Eq.~(\ref{eq:heatft})     clearly    resembles    the     usual    FT,
Eq.~(\ref{eq:ft}).  Indeed, if we invoke macroscopic thermodynamics to
argue that $-Q/T_A$ is the entropy change of $A$, and $+Q/T_B$ is that
of $B$,  then the  net entropy  generated by the  exchange of  heat is
given  by $\Sigma  = \Delta\beta  \cdot Q$,  and Eq.~(\ref{eq:heatft})
becomes Eq.~(\ref{eq:ft}).   However, this argument works  only if the
heat transferred is very small  in comparison with the internal energy
of  either body,  whereas the  validity of  Eq.~(\ref{eq:heatft}) does
{\it  not}   require  this  assumption.   Therefore,   we  will  leave
Eq.~(\ref{eq:heatft}) as a  statistical statement about heat exchange,
rather  than  trying to  force  it to  be  a  statement about  entropy
generation {\it per se}.

To  derive Eq.~(\ref{eq:heatft}) from  classical equations  of motion,
let  ${\bf z}_A$  denote the  phase space  coordinates  specifying the
microstate of  body $A$  (e.g.\ the positions  and momenta of  all its
degrees of freedom);  and let $H^A({\bf z}_A)$ be  a Hamiltonian whose
value  defines  the internal  energy  of $A$,  as  a  function of  its
microstate.   Similarly for $H^B({\bf  z}_B)$.  Let  $h^{\rm int}({\bf
  z}_A,{\bf z}_B)$  denote a small interaction term,  turned ``on'' at
$t=0$, and  ``off'' at $t=\tau$,  coupling the two bodies.   Let ${\bf
  y}=({\bf z}_A,{\bf  z}_B)$ specify a  point in the {\it  full} phase
space of all participating degrees of freedom.  During any realization
of the process  in which we are interested,  the microscopic evolution
of the two bodies is  described by a trajectory ${\bf y}(t)$, evolving
from $t=0$ to $t=\tau$ under Hamilton's equations, as derived from the
Hamiltonian
\begin{equation}
  {\cal H}({\bf y}) = H^A({\bf z}_A) + H^B({\bf z}_B) 
  + h^{\rm int}({\bf y}).
\end{equation}

\begin{figure}[htbp]
  \centering
  \includegraphics[scale=0.5]{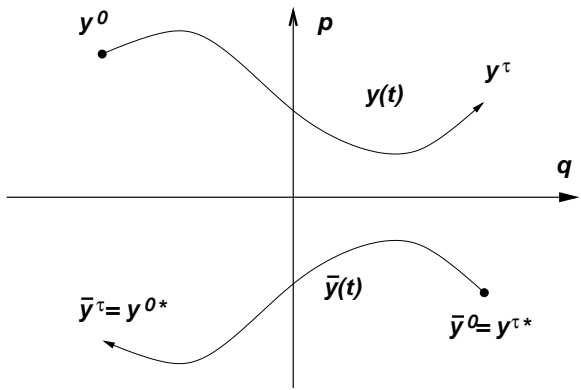}
  \caption{Twin trajectories $y(t)$ and $\bar{y}(t) = y^*(\tau -t)$
    related by time reversal.} 
  \label{fig:twins}
\end{figure}

We now further assume {\it time-reversal invariance}:
\begin{equation}
  \label{eq:tri}
  H^i({\bf z}_i) = H^i({\bf z}_i^*) ,\quad
  h^{\rm int}({\bf y}) = h^{\rm int}({\bf y}^*),
\end{equation}
where  $i=A,B$   and  the  asterisk  (*)   denotes  the  time-reversal
operation,  usually the reversal  of momenta:  $({\bf q},{\bf  p})^* =
({\bf q},-{\bf p})$.  This assumption has the crucial consequence that
microscopic  realizations of  the  process come  in  pairs related  by
time-reversal: for any trajectory ${\bf  y}(t)$ which is a solution of
Hamilton's equations,  its time-reversed twin,  $\overline{\bf y}(t) =
{\bf  y}^*(\tau-t)$, is  also a  solution.  For  future  reference let
${\bf y}^0$ and ${\bf y}^\tau$ denote the initial and final conditions
of the  ``forward'' realization~\footnote{ Of course, for  any pair of
  realizations related  by time-reversal, the designation  of which is
  the forward  realization, and which is the  reverse, is arbitrary.},
${\bf  y}(t)$;  hence   the  ``reverse''  realization,  $\overline{\bf
  y}(t)$,  evolves  from  $\overline{\bf  y}^0={\bf  y}^{\tau  *}$  to
$\overline{\bf    y}^\tau={\bf    y}^{0*}$,    as    illustrated    in
Figure~\ref{fig:twins}.

By  our assumption regarding  the equilibrium  preparation of  the two
bodies, the  probability distribution for  sampling initial conditions
${\bf y}^0$ is given by:
\begin{equation}
  \label{eq:heatinitprobdist}
  P({\bf y}^0) =
  {1\over Z_AZ_B}
  e^{-H^A({\bf z}_A^0)/T_A}
  e^{-H^B({\bf z}_B^0)/T_B},
\end{equation}
where  the $Z$'s are  partition functions.   Given a  trajectory ${\bf
  y}(t)$ and  its time-reversed twin $\overline{\bf  y}(t)$, the ratio
of probabilities  of sampling  their respective initial  conditions is
then:
\begin{equation}
  {P({\bf y}^0)\over P(\overline{\bf y}^0)} =
  e^{\Delta E_B/T_B} e^{\Delta E_A/T_A},
\end{equation}
where  $  \Delta  E_A  =  H^A({\bf z}_A^\tau)  -  H^A({\bf  z}_A^0)  =
H^A(\overline{\bf  z}_A^0) -  H^A({\bf  z}_A^0) $,  and similarly  for
$\Delta E_B$.  The quantities  $\Delta E_A$ and $\Delta E_B$ represent
the net  change in the internal  energies of the two  bodies, over the
course of  the realization described  by ${\bf y}(t)$.  If  we neglect
the  small  amount of  work  performed in  switching  on  and off  the
interaction term $h^{\rm  int}$, then the net change  in the energy of
one system is  compensated by an opposite change in  the energy of the
other, i.  e.  $\Delta E_B \approx  -\Delta E_A$, and it is natural to
view these changes as representing  a quantity of heat transfered from
$A$ to $B$: $Q := \Delta E_B \approx -\Delta E_A$.  Hence,
\begin{equation}
  \label{eq:heatratio}
  { P({\bf y}^0)\over P(\overline{\bf y}^0)} =
  e^{\Delta\beta\cdot \hat{Q}({\bf y}^0)},
\end{equation}
where the function $\hat Q({\bf y})$ denotes the value of $Q$ during a
realization evolving from initial conditions ${\bf y}$.  Note that
\begin{equation}
  \hat Q(\overline{\bf y}^0) = -\hat Q({\bf y}^0),
  \label{eq:heatopp}
\end{equation}
that is, the heat transfer  during the forward realization is opposite
to that during the reverse realization.

Combining Eqs.~(\ref{eq:heatratio}) and~(\ref{eq:heatopp}) we get:
\begin{eqnarray}
  p_\tau(Q) &=& \int d{\bf y}^0
  P({\bf y}^0) \delta[Q-\hat Q({\bf y}^0)] \nonumber \\
  &=& e^{\Delta\beta\cdot Q}
  \int d\overline{\bf y}^0
  P(\overline{\bf y}^0)
  \delta[Q + \hat Q(\overline{\bf y}^0)] \nonumber \\ 
  &=&
  e^{\Delta\beta\cdot Q} 
  p_\tau(-Q),\label{eq:advertised} 
\end{eqnarray}
which is equivalent to  Eq.~(\ref{eq:heatft}).  Here the change in the
variables  of  integration  between  the  first and  second  lines  is
justified  by  the invariance  of  the  Liouville  measure under  time
evolution  ($d{\bf y}^0  =  d{\bf  y}^\tau$), as  well  as under  time
reversal ($d{\bf y}^\tau = d{\bf y}^{\tau*} = d{\overline{\bf y}}^0$).

These formal manipulations can be understood intuitively.  $p_\tau(Q)$
is a  sum of  contributions from all  realizations for which  the heat
transfer takes  on a specified value,  $Q$; and $p_\tau(-Q)$  is a sum
over those for which the heat transfer is $-Q$.  But these two sets of
realizations  are in one-to-one  correspondence; for  every trajectory
${\bf  y}(t)$ belonging  to the  former set,  its  twin $\overline{\bf
  y}(t)$ belongs to  the latter (Eq.~\ref{eq:heatopp}).  Moreover, the
ratio of  initial condition sampling probabilities for  such a twinned
pair      of     realizations     is      $e^{\Delta\beta\cdot     Q}$
(Eq.~(\ref{eq:heatratio})).   Therefore,  when  we  add  the  sampling
probabilities $P({\bf y}^0)$ from the first set of realizations to get
$p_\tau(Q)$, and  $P(\overline{\bf y}^0)$ from  the second set  to get
$p_\tau(-Q)$, the ratio of the sums is $e^{\Delta\beta\cdot Q}$.

The above  derivation, based  on comparing the  sampling probabilities
for pairs of  twinned trajectories, is similar to  that carried out by
Evans  and Searles~\cite{EvansS94}  for the  transient FT.   Note also
that this derivation is valid for arbitrary times $\tau$; there are no
hidden  assumptions that the  temperatures of  the two  systems remain
constant, or even well-defined after $t=0$.

The  sole  approximation that  we  have made  is  the  neglect of  the
interaction term, $h^{\rm int}$.  In  reality, a finite amount of work
is  required to  turn on  this interaction  initially,  $\delta w_{\rm
  on}$, and  then to turn it  off finally, $\delta  w_{\rm off}$.  The
resulting balance of  energy reads: $\Delta E_A +  \Delta E_B = \delta
w_{\rm on} + \delta w_{\rm off}$,  hence $\delta w = \delta w_{\rm on}
+  \delta w_{\rm  off}$ enters  as a  correction to  the approximation
$\Delta E_B  \approx -\Delta E_A$  used earlier.  The validity  of our
final result  thus requires  that the work  performed in  coupling and
later uncoupling the  systems ($|\delta w|$) be much  smaller than the
typical  energy  change in  either  system  ($|\Delta E_A|$,  $|\Delta
E_B|$).  Whether or  not this condition is met  depends, of course, on
details of  the two systems, on  the strength of  the interaction term
($\delta w \sim h^{\rm int}$), and on the duration $\tau$.  It will be
interesting  to investigate  this  issue in  the  context of  specific
models.

We proceed now to the proof  of the quantum version of our theorem. We
assume  that systems  $A$ and  $B$ have  equilibrated  to temperatures
$T_A, T_B$  before the experiment,  and are thus described  by density
matrices $\rho_i  = \exp(- \beta_i H^i)/Z_i$, where  $i=A,B$.  At time
$t  = 0^-$ we  separate the  systems from  the reservoirs  and measure
their energies.  As a result, each system $i$ is projected onto a pure
state $|n_i\rangle$ with probability $e^{-\beta_i E_{n_i}^i}/Z_i$, and
the  combined   system  is  described  by  the   product  state  $|n_A
n_B\rangle$.  We  then allow  the systems to  interact through  a weak
coupling term $h^{\mathrm{int}}$.  Thus the Hamiltonian takes the form
${\cal H} = H^A\otimes I^B + I^A \otimes H^B + h^{\mathrm{int}}$.

Let us now assume, as in the classical case (Eq.~(\ref{eq:tri})), that
the system  and both its  subsystems are time-reversal  invariant.  In
quantum  mechanics  the  time-reversal   invariance  of  a  system  is
expressed by the condition
\begin{equation}
  \label{eq:qtri}
  \Theta H = H \Theta,
\end{equation}
where  $H$ is  the system  Hamiltonian,  and $\Theta$  is the  quantum
time-reversal     operator~\cite{Merzbacher98,Ballentine98}.      This
operator reverses  linear and angular momentum  while keeping position
unchanged, and is {\it anti-linear}:
\begin{equation}
  \Theta \Bigl( \alpha_1 |\psi\rangle + \alpha_2 |\phi\rangle \Bigr)
  = \alpha_1^\dagger \Theta |\psi\rangle +
    \alpha_2^\dagger \Theta |\phi\rangle,
\end{equation}
where the dagger denotes  complex conjugation.  When dealing with such
operators, the  Dirac bra-ket notation,  invented to deal  with linear
operators,       becomes        cumbersome:       the       expression
$\langle\phi|\Theta|\psi\rangle$ is ambiguous until we specify whether
$\Theta$  is acting  to  the right  or  to the  left.   To avoid  this
inconvenience  we  will use  the  standard  product  in Hilbert  space
$(|\phi\rangle ,  | \psi \rangle)$,  rather than the  more abbreviated
Dirac  bra-ket, $\langle  \phi |  \psi \rangle$,  to denote  the inner
product  between  two  wave  functions.  From  Eq.~(\ref{eq:qtri})  it
follows  that,   for  every   eigenstate  $|n\rangle$  of   $H$  there
corresponds a  time-reversed eigenstate  $\Theta |n \rangle$  with the
same energy; these two states are either linearly independent, or else
identical  apart  from an  overall  phase.   Moreover, since  $\Theta$
preserves wave function normalization,  it is not just anti-linear but
also {\it anti-unitary}: $(  \Theta |\phi\rangle , \Theta |\psi\rangle
)  = (  |\psi\rangle ,  |\phi\rangle )$.   We will  make use  of these
properties in the analysis below.

Having turned on  the interaction term at $t=0$,  we allow the systems
to evolve for a time $\tau$.  The combined system then reaches a state
$|\Psi\rangle$, obtained  from the initial state  $|n_A n_B\rangle$ by
evolution under  Schr\" odinger's equation.   We now separate  the two
systems -- that is, we turn off the interaction term -- and once again
measure their  energies.  The  state $|\Psi\rangle$ is  thus projected
onto  a  product state  $|m_A  m_B\rangle$.   As  before, we  make  no
assumptions  regarding  $\tau$, in  particular  the  systems have  not
necessarily equilibrated.

Letting  $P_\tau(|n\rangle  \rightarrow   |  m  \rangle)$  denote  the
probability of observing  a transition from $| n  \rangle \equiv | n_A
n_B \rangle$ to $| m \rangle \equiv | m_A m_B \rangle$, we have
\[
P_\tau(|n\rangle  \rightarrow | m \rangle)  =  |( | m \rangle  ,
U_\tau  | n \rangle)  |^2 \frac{ e^{-\beta_A  E_{n_A}^A -\beta_B
    E_{n_B}^B }}{Z_A   Z_B} ,
\]
where  $U_\tau =  e^{-  i \tau  {\cal  H}}$ is  the quantum  evolution
operator, and  $\hbar =  1$.  The  second factor on  the right  is the
probability for  sampling the initial  state $| n \rangle$;  the first
factor is the transition probability from $|n\rangle$ to $| m\rangle$.
Similarly, the  probability of observing  the time-reversed transition
from $ \Theta | m \rangle $ to $ \Theta | n\rangle $ is
\[
\!\! P_\tau(\Theta |m \rangle \rightarrow \Theta |n\rangle) = |(
\Theta |n  \rangle , U_\tau  \Theta| m \rangle)  |^2 \frac{e^{-\beta_A
    E_{m_A}^A -\beta_B E_{m_B}^B }}{Z_A Z_B}.
\]
But,  since $\Theta$  is  anti-unitary, and  $U_\tau  \Theta =  \Theta
U_{-\tau}$~\footnote{  This  follows  from  Eq.\ref{eq:qtri}  and  the
  anti-linearity of $\Theta$.}, we have
\begin{eqnarray*}
  ( \Theta |n \rangle , U_\tau \Theta| m \rangle) & = &
  ( \Theta |n \rangle , \Theta U_{-\tau}| m \rangle)
  = ( U_{-\tau}| m \rangle, |n\rangle) \\
  & = &  ( | m \rangle, U_\tau|n\rangle),
\end{eqnarray*}
therefore 
\begin{equation}
  \frac{
    P_\tau(|n\rangle \rightarrow |m\rangle )
  }{
    P_\tau(\Theta |m\rangle  \rightarrow \Theta |n\rangle ) 
  } = e^{-\beta_A (E_{n_A}^A -E_{m_A}^A)} e^{-\beta_B (E_{n_B}^B -
    E_{m_B}^B )}.
\end{equation}
Since we assumed that the interaction is weak, we expect the energy of
the total system to be almost preserved:
\begin{equation}
  E_n^A + E_n^B \approx E_m^A + E_m^B.
\end{equation}
It  follows   that  the  energy   changes  in  the  two   systems  are
approximately equal
\begin{equation}
  \label{eq:12}
  Q_{n \rightarrow m} := E_m^B - E_n^B \approx E_n^A - E_m^A.
\end{equation}
We interpret $Q$ as the heat exchange between the systems $A$ and $B$.
Thus,
\begin{equation}
  \frac{
    P_\tau(|n\rangle \rightarrow |m\rangle )
  }{
    P_\tau(\Theta |m\rangle  \rightarrow \Theta |n\rangle  ) 
  } \approx e^{\Delta \beta \cdot  Q_{n \rightarrow m}}.
\end{equation}
Since every eigenstate has a corresponding time-reversed twin, the net
probability of the heat transfer $Q$ in time $\tau$ is
\begin{eqnarray}
  p_\tau(Q) & = & \sum_{n,m} P_\tau(|n\rangle \rightarrow |m\rangle)
    \delta(Q - Q_{n \rightarrow m}) \nonumber \\
  & = & e^{\Delta \beta \cdot Q} \sum_{\Theta n,\Theta m}
    P_\tau(\Theta |m\rangle \rightarrow \Theta |n\rangle ) \delta(Q +
    Q_{\Theta m  
    \rightarrow \Theta n}) \nonumber \\
  & = & e^{\Delta \beta \cdot  Q} p_\tau(-Q). \label{eq:13.5}
\end{eqnarray}
This result  is true for the  quantities as we have  defined them.  We
can rewrite  Eq.~(\ref{eq:13.5}) in the  form of Eq.~(\ref{eq:heatft})
if  we   further  assume  a  sufficiently  dense   spectrum,  so  that
$p_\tau(Q)$ can be replaced by a locally smooth function.

At  the level  of macroscopic  thermodynamics (and  in the  absence of
external work),  the passage of  heat from a  colder to a  hotter body
constitutes    a    violation     of    the    Second    Law.     From
Eq.~(\ref{eq:heatft}), we can derive an upper bound on the probability
of observing such a ``violation'',  of at least some finite magnitude,
as  follows.   Assume  that  $T_A>T_B$,  i.e.\  $\Delta\beta>0$.   The
probability that the  heat transfer from $A$ to $B$  will fall below a
specified  value $q$  is  given by  $\int_{-\infty}^q  p_\tau(Q) dQ$.  
Using     Eq.~(\ref{eq:heatft})    to    replace     $p_\tau(Q)$    by
$p_\tau(-Q)\exp(\Delta\beta\cdot Q)$, and then invoking the inequality
chain
\[
  \int_{-\infty}^q p_\tau(-Q)
  e^{\Delta\beta\cdot Q} dQ
  \le
  e^{\Delta\beta\cdot q}
  \int_{-\infty}^q p_\tau(-Q) dQ
  \le
  e^{\Delta\beta\cdot q},
\]
we get
\begin{equation}
  \label{eq:bound}
  \int_{-\infty}^q p_\tau(Q)\, dQ
  \le e^{\Delta\beta\cdot q}.
\end{equation}
Choosing $q<0$, this result tells us that the probability of observing
a  net heat  transfer in  the  ``wrong'' direction  ($Q<0$), from  $B$
(cold) to $A$  (hot), of at least some  magnitude $\vert q\vert$, dies
exponentially (or faster)  with that magnitude.  Eq.~(\ref{eq:heatft})
also implies that the average of $\exp(-\Delta\beta\cdot Q)$, over the
ensemble of realizations for any time $\tau$, is unity:
\begin{equation}
  \overline{e^{-\Delta\beta\cdot Q}}
  \equiv \int dQ p_\tau(Q) e^{-\Delta\beta\cdot Q} = 1.
\end{equation}

In conclusion,  a result  analogous to the  FT for  entropy generation
(Eq.~(\ref{eq:ft})), and  valid for  arbitrary times $\tau$,  has been
derived for  the statistics of heat exchange  between finite classical
or    quantum    systems    separately   prepared    in    equilibrium
(Eqs.~(\ref{eq:heatft})).   In our  derivation  we invoke  statistical
mechanics  to describe the  initial preparation  of the  systems, then
treat their evolution during  the interval of contact dynamically.  We
also  assume  a  negligible  energy  of interaction  between  the  two
systems, and  a time-reversal  invariant Hamiltonian.  In  the quantum
case, an additional source of randomness arises from the fact that the
initial quantum  state of the  system does not uniquely  determine the
outcome of the final energy measurements.  Nevertheless, this does not
spoil our  result.  We finally mention  that a similar  theorem can be
derived  for particle  exchange between  two reservoirs,  driven  by a
difference in initial chemical potentials (unpublished).

It is  a pleasure to  thank J.R.  Anglin,  H.  van Beijeren, and  D.J. 
Thouless for stimulating and  useful conversations.  This research was
supported by  the Department of Energy,  under contract W-7405-ENG-36,
and  by the Polish-American  Maria Sk\l  odowska-Curie Joint  Fund II,
under project  PAA / DOE-98-343.   DW thanks the Center  for Nonlinear
Science,  School  of  Physics,  Georgia Institute  of  Technology  for
support as a Joseph Ford Fellow.

\end{document}